\documentclass[aps, a4paper, twocolumn, superscriptaddress, amsmath]{revtex4}

\usepackage{epsf, latexsym, color, epsfig}
\usepackage{amssymb, amsmath}
\usepackage{times}
\usepackage{qcircuit}
\usepackage{verbatim}
\usepackage{booktabs}
\usepackage{multirow}

\usepackage[colorlinks]{hyperref}
\AtBeginDocument{%
    \hypersetup{
      urlcolor    = blue, 
      filecolor   = red,
      linkcolor   = blue, 
      citecolor   = blue, 
    }
} 

\newcommand{\bra}[1]{\langle #1 |}
\newcommand{\ket}[1]{| #1 \rangle}

\newcommand\h{{\cal H}}

\newcommand{\etal}{{\textit{et al.}}}

\newcommand\tr{{\mbox{Tr\,}}}

\newcommand{\ignore}[1]{}

\newcommand{\be}{\begin{equation}}
\newcommand{\ee}{\end{equation}}
\newcommand{\ba}{\begin{eqnarray}}
\newcommand{\ea}{\end{eqnarray}}
\newcommand{\bc}{\begin{center}}
\newcommand{\ec}{\end{center}}

\def\CC{{\rm\kern.24em \vrule width.04em height1.46ex depth-.07ex
    \kern-.30em C}}
\def\P{{\rm I\kern-.25em P}}

\def\RR{{\rm
         \vrule width.04em height1.58ex depth-.0ex
         \kern-.04em R}}

\def\bbbc{{\mathchoice {\setbox0=\hbox{$\displaystyle\rm C$}\hbox{\hbox
to0pt{\kern0.4\wd0\vrule height0.9\ht0\hss}\box0}}
{\setbox0=\hbox{$\textstyle\rm C$}\hbox{\hbox
to0pt{\kern0.4\wd0\vrule height0.9\ht0\hss}\box0}}
{\setbox0=\hbox{$\scriptstyle\rm C$}\hbox{\hbox
to0pt{\kern0.4\wd0\vrule height0.9\ht0\hss}\box0}}
{\setbox0=\hbox{$\scriptscriptstyle\rm C$}\hbox{\hbox
to0pt{\kern0.4\wd0\vrule height0.9\ht0\hss}\box0}}}}

\def\bbbq{{\mathchoice {\setbox0=\hbox{$\displaystyle\rm Q$}\hbox{\raise
0.15\ht0\hbox to0pt{\kern0.4\wd0\vrule height0.8\ht0\hss}\box0}}
{\setbox0=\hbox{$\textstyle\rm Q$}\hbox{\raise
0.15\ht0\hbox to0pt{\kern0.4\wd0\vrule height0.8\ht0\hss}\box0}}
{\setbox0=\hbox{$\scriptstyle\rm Q$}\hbox{\raise
0.15\ht0\hbox to0pt{\kern0.4\wd0\vrule height0.7\ht0\hss}\box0}}
{\setbox0=\hbox{$\scriptscriptstyle\rm Q$}\hbox{\raise
0.15\ht0\hbox to0pt{\kern0.4\wd0\vrule height0.7\ht0\hss}\box0}}}}

\def\bbbt{{\mathchoice {\setbox0=\hbox{$\displaystyle\rm
T$}\hbox{\hbox to0pt{\kern0.3\wd0\vrule height0.9\ht0\hss}\box0}}
{\setbox0=\hbox{$\textstyle\rm T$}\hbox{\hbox
to0pt{\kern0.3\wd0\vrule height0.9\ht0\hss}\box0}}
{\setbox6=\hbox{$\scriptstyle\rm T$}\hbox{\hbox
to0pt{\kern8.3\wd0\vrule height0.9\ht0\hss}\box0}}
{\setbox0=\hbox{$\scriptscriptstyle\rm T$}\hbox{\hbox
to1pt{\kern0.3\wd1\vrule height0.9\ht0\hss}\box0}}}}

\def\bbbz{{\mathchoice {\hbox{$\sf\textstyle Z\kern-0.4em Z$}}
{\hbox{$\sf\textstyle Z\kern-0.4em Z$}}
{\hbox{$\sf\scriptstyle Z\kern-0.3em Z$}}
{\hbox{$\sf\scriptscriptstyle Z\kern-0.2em Z$}}}}

\newcommand{\putfig}[2]{$$\leavevmode\hbox{\epsfxsize=#2 cm
   \epsffile{#1}}$$}

\begin{document}

\title{OAM tomography with Heisenberg-Weyl observables}

\author{Alexandra Maria P\u alici}
\altaffiliation{These authors contributed equally}
\affiliation{Horia Hulubei National Institute of Physics and Nuclear Engineering, 077125 Bucharest--M\u agurele, Romania}

\author{Tudor-Alexandru Isdrail\u a}
\altaffiliation{These authors contributed equally}
\affiliation{Horia Hulubei National Institute of Physics and Nuclear Engineering, 077125 Bucharest--M\u agurele, Romania}

\author{Stefan Ataman}
\affiliation{Extreme Light Infrastructure-Nuclear Physics (ELI-NP), Horia Hulubei National Institute for Physics and Nuclear Engineering, 30 Reactorului Street, 077125 Bucharest--M\u agurele, Romania}

\author{Radu Ionicioiu}
\affiliation{Horia Hulubei National Institute of Physics and Nuclear Engineering, 077125 Bucharest--M\u agurele, Romania}

\begin{abstract}
Photons carrying orbital angular momentum (OAM) are excellent qudits and are widely used in several applications, such as long distance quantum communication, $d$-dimensional teleportation and high-resolution imaging and metrology. All these protocols rely on quantum tomography to characterise the OAM state, which currently requires complex measurements involving spatial light modulators and mode filters. To simplify the measurement and characterisation of OAM states, here we apply a recent tomography protocol [Asadian \etal, \pra {\bf 94}, 010301 (2016)]. Our scheme for OAM tomography in $d$ dimensions requires only a set of measurements on a mode qubit, i.e., a 2-dimensional system. This replaces the current complexity of OAM measurements by the ability to perform generalized Pauli operators $X_d, Z_d$ on OAM states. Our scheme can be adapted in principle to other degrees of freedom, thus opening the way for more complex qudit tomography.
\end{abstract}

\maketitle

\section{Introduction}

Quantum state tomography (QST), or quantum state reconstruction, is a technique used to extract the maximum available information from a quantum state in order to reconstruct its density matrix (for discrete variables) or its phase-space representation (for continuous variables) \cite{Fano1957}. 

Quantum tomography has been performed experimentally on various physical systems, e.g., sub-levels of H$^+$ and He atoms \cite{Ashburn1990}, coherent states \cite{Smithey1993b}, squeezed vacuum states \cite{Smithey1993, Schiller1996, Breitenbach1997}, vibrations of molecules \cite{Leonhardt1997} and large angular momentum states of Cs atoms \cite{Klose2001}.

The simplest QST, on polarisation qubits, consists of measuring the well-known Stokes parameters \cite{Stokes1852}. This technique was demonstrated on a non-maximally entangled bi-photon state \cite{White1999}. It can be extended to $n$ qubits \cite{james} and was experimentally demonstrated for two qubits \cite{james}.

Scaling QST from qubits to qudits is challenging for several reasons. First, one has to find the equivalent of Pauli operators in $d$ dimensions. One popular choice are the Gell-Mann matrices $\lambda_i$, the generators of $\mathrm{SU}(d)$. Second, in order to reconstruct $\rho$ we need to measure experimentally the expectations values $\langle \lambda_i \rangle$, which is not trivial \cite{thew}.

A different approach to qudit tomography has been introduced recently by Asadian \etal~\cite{HWeyl}. The idea is to use the Heisenberg-Weyl observables (HW) instead of $\lambda_i$. In the same article \cite{HWeyl} the authors also proposed a method to measure HW observables based on deterministic quantum computation with one qubit (DQC1) model \cite{DQC1}. This scheme is particularly attractive since it requires to measure only a qubit (ancilla) instead of measuring the qudit (see next section).

In this article we apply the Asadian \etal~scheme to an experimentally important case, namely photonic orbital angular momentum (OAM). Since the seminal paper of Allen \etal~\cite{Allen1992}, orbital angular momentum of light became a new field of exploration \cite{Molina-Terriza2001, Leach2002}. The orbital angular momentum of a photon is due to the helical phase front along the propagation direction with quantized angular momentum $\ell \hbar$, with $\ell \in \mathbb Z$. A review of OAM states can be found in Ref.~\cite{Molina-Terriza2007, Barnett2017, YShen}.

Due to their weak coupling to the environment, photons carrying OAM are a natural choice to implement protocols for qudits. Applications include QKD \cite{Boyd2011}, object identification \cite{Uribe-Patarroyo2013}, enhanced phase sensitivity \cite{DAmbrosio2013}, imaging with super-diffraction-limit resolution \cite{Tamburini2006} and metrology \cite{Lavery2013, Cvijetic2015}.

All these applications rely on quantum tomography to characterise the OAM state. Current methods for OAM tomography \cite{Leach2002, Nicolas2015, Jack2009, agnew} require fork holograms, usually displayed by spatial light modulators (SLMs), and single-mode fibers, acting as mode filters. SLMs have low efficiency due to the pixelated surface and the existence of dead zones (the opaque areas between pixels). Mode filters further reduce the efficiency.

Here we describe a different setup for OAM tomography. Our protocol simplifies the measurement step, as we need to measure only a mode qubit. In our case the complexity is shifted to the ability to apply generalised Pauli gates on OAM states.

The article is structured as follows. In Section \ref{sec:background} we briefly describe qudit tomography with Heisenberg-Weyl observables. In Section \ref{sec:proposal} we apply this method to quantum tomography for OAM states. Our scheme uses a Mach-Zehnder interferometer with tuneable phases and generalised Pauli gates $Z_d^l X_d^m$. Finally, we sketch future perspectives of our setup in Section \ref{sec:conclusions}.

\section{Quantum tomography in $d$-dimensions}
\label{sec:background}

Performing tomography of the density matrix $\rho$ requires a suitable basis. In the following we assume the Hilbert space $\h$ is finite dimensional, $\dim \h=d$. For example, in $d=2$ we use the Pauli matrices $\{ X_i, i=0,\ldots, 3\}:= \{\mathbb{I}_2, X, Y, Z \}$, such that:
\be
\rho= \frac{1}{2} \sum_{i=0}^3 r_i X_i
\ee
The Pauli matrices are Hermitian and form a basis in the space ${\cal M}(2)$ of linear operators in 2-dimensions. The coefficients $r_i= \tr(\rho X_i)= \langle X_i \rangle$ are the expectation values of the Pauli operators on the state $\rho$:
\be
\rho= \frac{1}{2} \begin{bmatrix} 1+ \langle Z \rangle & \langle X \rangle- i \langle Y \rangle \cr \langle X \rangle+ i \langle Y \rangle & 1- \langle Z \rangle \end{bmatrix}
\ee
In $d$ dimensions a possible basis is the set of generalised Gell-Mann matrices $\lambda_i$ \cite{thew}, the standard generators of $\mathrm{SU}(d)$. However, it is not straightforward to measure the expectation values $\langle \lambda_i \rangle$ of these observables in $d$ dimensions.

Here we adopt a different approach and use the Heisenberg-Weyl (HW) observables introduced in Ref.~\cite{HWeyl}. Consider the generalised Pauli operators $X_d$ and $Z_d$ in $d$ dimensions: 
\ba
\label{Xd}
X_d:\ \ &&\ket{j} \mapsto \ket{j\oplus 1}\\
\label{Zd}
Z_d:\ \ &&\ket{j} \mapsto \omega^j \ket{j}
\ea
where $\oplus$ is addition mod $d$ and $\omega= e^{2\pi i/d}$ a root of unity of order $d$. These operators are unitary but not Hermitian, $X_d^\dag= X_d^{-1}$, $Z_d^\dag= Z_d^{-1}$, $X_d^d= \mathbb{I}_d= Z_d^d$; therefore they are not observables. Nevertheless, we can use them to construct the Heisenberg-Weyl (HW) observables \cite{HWeyl}:
\be
\label{eq:Q_lm_definition_HW}
Q_{lm}= \frac{1+i}{2} Z_d^l X_d^m \omega^{-lm/2}+ h. c.
\ee
The operators $Q_{lm}$ are Hermitian (by construction) and orthogonal \cite{HWeyl}:
\be
\tr\{Q_{lm} Q_{l' m'} \}= d \delta_{l l'} \delta_{m m'}
\ee
This gives us a set of $d^2$ linearly independent observables which form a basis in the space of $d$-dimensional linear operators ${\cal M}(d)$. Therefore any density operator can be written as: 
\be
\rho= \frac{1}{d} \sum_{l, m=0}^{d-1} \langle Q_{lm} \rangle Q_{lm}
\ee
with
\be
\langle Q_{lm} \rangle= \tr(\rho Q_{lm})
\ee
Thus tomography of a $d$-dimensional density matrix $\rho$ reduces to the measurement of the expectation values $\langle Q_{lm} \rangle$. Asadian \etal~\cite{HWeyl} showed how to measure these values using the DQC1 technique. Deterministic quantum computation with one qubit is an efficient method to estimate the normalised trace of an operator \cite{DQC1}. We briefly discuss this method in the following.

\begin{figure}
	\putfig{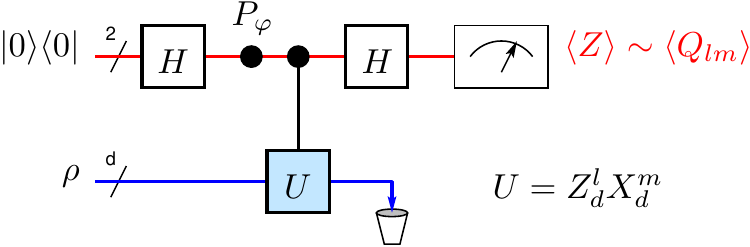}{8.5}
	\caption{Measuring the trace of an operator $U$ in DQC1. A qubit ancilla (red) is coupled to a qudit (blue) via a $C(U)$ gate. Measuring the qubit gives us information about the trace in the qudit subspace $\tr\{ \rho e^{i\varphi} U + h.c.\}$, eq.~\eqref{ev_Z}.}
	\label{fig:circuit}
\end{figure}

Consider the circuit in Fig.\ref{fig:circuit}. We use a 2-dimensional ancilla (a qubit) to perform tomography of a $d$-dimensional density matrix $\rho$ of a qudit. The initial state of the qubit-qudit system is separable:
\be
\label{eq:rho_in_global}
\rho_{in}={\vert0\rangle\langle0\vert}\otimes\rho
\ee
On the qubit ancilla we perform a Hadamard $H$, a phase shift $P_\varphi$, then a controlled unitary $C(U)$ that couples the qubit and the qudit. Finally, we perform another Hadamard $H$ and we then measure the qubit in the $Z$-basis (Fig.~\ref{fig:circuit}). After a straightforward calculation we obtain the expectation value of the qubit, see Appendix \ref{sec:app:calculations_rho}:
\be
\langle Z \rangle= \tfrac{1}{2} \tr\{ \rho (e^{i\varphi} U + e^{-i\varphi} U^\dag)\}
\label{ev_Z}
\ee
Importantly, no measurement is perform on the qudit, as we trace out this system. Since the $C(U)$ gate provides an effective qubit-qudit coupling (i.e., it entangles the two subsystems), a measurement on the qubit gives us information about the qudit state.

Choosing $U= Z_d^l X_d^m$ and taking $\varphi= \varphi_{lm}:= \tfrac{\pi}{4}- \tfrac{\pi}{d}lm$, we have $Q_{lm}= \tfrac{1}{\sqrt 2} e^{i\varphi_{lm}} U + h.c.$, therefore
\be
\langle Q_{lm}\rangle = \tr(\rho Q_{lm})= \sqrt{2} \langle Z \rangle
\ee
Thus we can perform tomography of a $d$-dimensional density matrix $\rho$ only by measuring the expectation value $\langle Z \rangle$ of a qubit. All the coefficients $\langle Q_{lm} \rangle$ are obtained by changing the phases $\varphi_{lm}$ and the controlled operator $C(Z_d^l X_d^m)$, $l,m=0,\ldots, d-1$, between the ancilla and the qudit.

\section{Quantum tomography for OAM}
\label{sec:proposal}

We now apply the previous method to photons carrying orbital angular momentum, a popular choice for qudits. This requires the following ingredients:\\
(i) a fully-controllable qubit ancilla which can be measured in the $Z$-basis; fully-controllable means we can apply $H$ and $P_{\varphi_{lm}}$ gates;\\
(ii) the capability to apply a controlled gate $C(U)$ between qubit and OAM qudit;\\
(iii) the capability to implement $Z_d^l X_d^m$ operators on the OAM qudit, $l,m=0,\ldots, d-1$.

In the following we discuss how to implement these requirements.

\begin{figure}
	\putfig{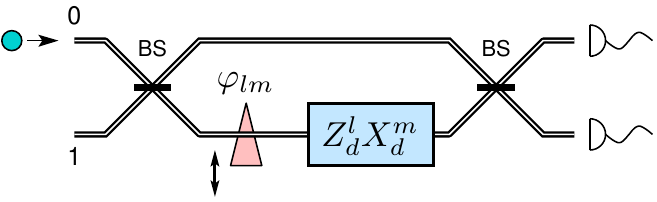}{8}
	\caption{Experimental setup for OAM qudit tomography using a Mach-Zehnder interferometer. The $P_{\varphi_{lm}}$ gate is implemented with an optical path difference in arm 1, independent of the OAM state. The controlled $C(Z_d^l X_d^m)$ gate is performed with unitaries $Z_d^l X_d^m$, acting on OAM, located only on mode-1 of the MZI.}
	\label{fig:circuit_implementation}
\end{figure}

\noindent (i) {\em A priori} there are several ways to implement the ancilla. The qubit can be a different quantum system (photon, atom in a cavity etc) or it can be another degree of freedom of the photon. Because we need to apply a controlled gate $C(U)$ and since it is experimentally difficult to implement a photon-photon or photon-atom interaction, the natural choice is to use the spatial mode of the photon as the qubit ancilla. In this case the Hadamard gate $H$ is a beamsplitter \cite{KLMAir, AdamiCerf, Qiang2018, Heilman}, and therefore the two $H$ gates define a Mach-Zehnder interferometer (MZI), as in Fig.~\ref{fig:circuit_implementation}.

The phase-gate $P_{\varphi_{lm}}$ on the mode qubit (Fig.~\ref{fig:circuit}) is equivalent to a phase shift $\varphi_{lm}$ in one arm of the MZI interferometer, independently on the OAM value. This can be done by having a variable path difference between the two arms of the MZI interferometer, Fig.~\ref{fig:circuit_implementation}.

\noindent (ii) Using a spatial mode as the ancilla also solves the second problem. In this case the controlled gate $C(U)$ on the OAM is particularly simple: we need to apply the $U$ gate on the OAM only on spatial mode 1.
 
\begin{figure}[t]
\putfig{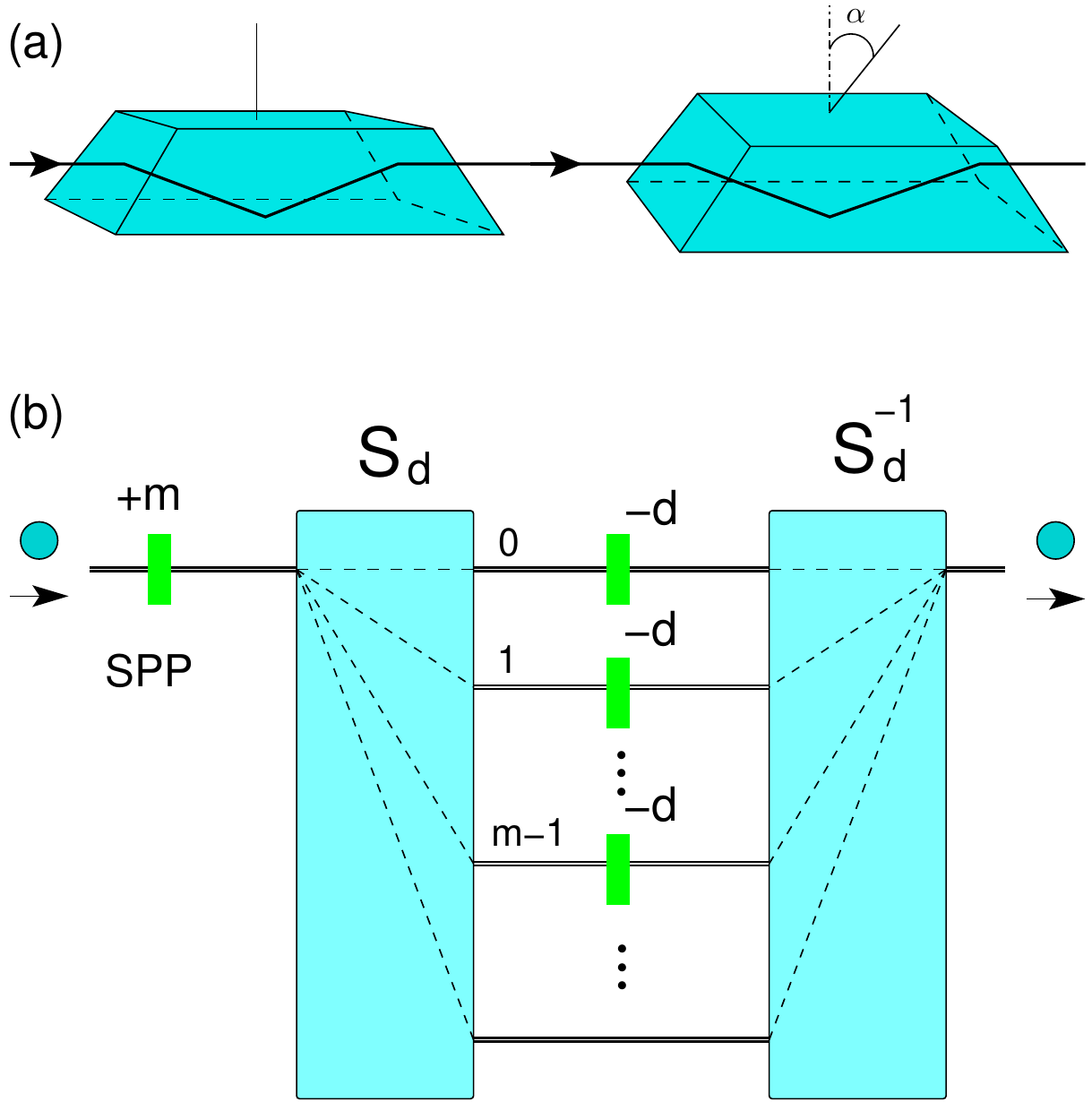}{6}
\caption{(a) $Z_d^l$ gate is implemented with a pair of Dove prisms rotated with a relative angle $\alpha = \pi l/d$. (b) $X_d^m$ gate (parallel setup). Inside the interferometer only the first $m$ modes have a spiral phase plate SPP$(-d)$.}
\label{ZlXm}
\end{figure}

\noindent (iii) Finally, we need to implement $Z_d^l X_d^m$ gates. For OAM states, the standard implementation of the $Z_d^l$ gate consists of two Dove prisms rotated with a relative angle $\alpha = \pi l/d$ \cite{Leach2002, Dove, Gonzalez2006}, Fig.~\ref{ZlXm}a
\be
Z_d^l:\ \ \ket{k} \rightarrow e^{i2\alpha k} \ket{k}= \omega^{lk} \ket{k}
\ee

The last building block is the $X_d$ gate which performs a cyclic permutation of the basis states \eqref{Xd}. Cyclic permutations for OAM states are difficult to implement and until recently only specific examples for $d=3,4$ were known \cite{Schlederer, DXChen}. Two methods to construct arbitrary $X_d$ gates for OAM states have been proposed recently \cite{Xd, Xd_gao}.

The first optical element of the $X_d$ gate is a spiral-phase plate (SPP), see Ref.~\cite{Xd}. A spiral-phase plate of order $k$, $\mathrm{SPP}(k)$, shifts all OAM values by $k$ units, $k\in \mathbb{Z}$:
\be
\mathrm{SPP}(k):\ \ \ket{j}_{OAM} \mapsto \ket{j+k}_{OAM}
\label{SPP}
\ee
When we apply $\mathrm{SPP}(1)$ to the basis states $\ket{j}$, the last state will be out of range, $\ket{d-1} \mapsto \ket{d}$; for simplicity, we omit the OAM subscript. For the cyclic permutation \eqref{Xd} we need to put this value back to $\ket{0}$.

This is where the OAM sorter $S_d$ \cite{sorter} comes into play. The sorter $S_d$ demultiplexes (i.e., sorts) different OAM values into distinct spatial modes and acts as a generalized polarizing beam-splitter (PBS) for OAM. Thus after applying $\mathrm{SPP}(1)$ and the sorter $S_d$, we apply $\mathrm{SPP}(-d)$ only to spatial mode 0. Finally, the inverse sorter $S_d^{-1}$ multiplexes back all OAM values on the same spatial mode (Fig.\ref{ZlXm}b).

In order to implement the $X_d^m$ gate there are two strategies. First, one could apply $m$ times the $X_d$ gate described above (the serial setup). However, this is rather inefficient. Here we propose a more efficient way (the parallel setup).

In the parallel version we first shift all states by $m$ units with a $\mathrm{SPP}(m)$. Consequently, $m$ OAM values will be out of range, $\ket{d},\ldots, \ket{d-1+m}$. Next a sorter $S_d$ demultiplexes all OAM values to different spatial modes. Since the sorter works cyclically on spatial modes \cite{sorter}, then the first $m$ spatial modes have OAM values outside of range. As before, in this case we shift back only these modes by $\mathrm{SPP}(-d)$. The last element is the inverse sorter $S_d^{-1}$ which multiplexes back all OAM states to a single spatial mode (i.e., it disentangles the mode and OAM degrees of freedom).

The setup for the parallel $X_d^m$ gate performs the following sequence (see Fig.~\ref{ZlXm}):
\ba
\nonumber \ket{i}_{OAM} \ket{0}_m &\stackrel{+m}{\longrightarrow}& \ket{i+m}_{OAM} \ket{0}_m \\
\nonumber &\stackrel{S_d}{\longrightarrow}& \ket{i+m}_{OAM} \ket{i\oplus m}_m \\
\nonumber &\stackrel{-d^{[r]}}{\longrightarrow}& \ket{i\oplus m}_{OAM} \ket{i\oplus m}_m \\
&\stackrel{S_d^{-1}}{\longrightarrow}& \ket{i\oplus m}_{OAM} \ket{0}_m
\ea
with $r=0,...,m-1$.

The resources required to implement $X_d^m$ in the serial vs. the parallel setup are summarised in Table \ref{table:resources}. Both cases use the same number of $\mathrm{SPP}(-d)$; the serial setup has $m$ $\mathrm{SPP}(1)$ and the parallel one has a single $\mathrm{SPP}(m)$ (which are equivalent resources). In contrast, the parallel setup needs a constant number of sorters (two) irrespective of $m$, whereas the serial setup requires $2m$ sorters $S_d, S_d^{-1}$. Thus the parallel setup saves $2m-2$ sorters.

\begin{table}[ht]
\centering
\begin{tabular}{l |c c c c}
 & \hspace{.1cm} $\mathrm{SPP}(1)$ & \hspace{.1cm} $\mathrm{SPP}(m)$ & \hspace{.1cm} $\mathrm{SPP}(-d)$ & \hspace{.1cm} $S_d, S_d^{-1}$ \\ [0.5ex]
\hline
serial $X_d^m$ & $m$ & 0 & $m$ & $2m$ \\ [1ex]
parallel $X_d^m$ & 0 & 1 & $m$ & 2 \\ [1ex]
\end{tabular}
\caption{Number of optical elements required to implement $X_d^m$ gate in the serial and parallel setup.}
\label{table:resources}
\end{table}

Resource-wise, the sorter $S_d$ requires two Fourier gates $F_d$ and $F_d^\dag$ acting on spatial modes and path-dependent phase shifts between them. The Fourier gates can be implemented using beam-splitters and phase-shifters \cite{Reck1994, Tabia2016, Clements2016} or as multimode interference devices in integrated optics \cite{JZhouFour, Cincotti12, Lowery10}.

\section{Discussion}
\label{sec:conclusions}

All protocols using photonic OAM as qudits require an efficient tomography step. Current methods for OAM tomography use spatial light modulators to display computer-generated holograms and single-mode fibres as mode filters \cite{Nicolas2015, Jack2009, agnew}. Although configurable, SLMs have a lower efficiency compared to SPPs. Moreover, they are bulky and difficult to miniaturize, given the current drive towards integrated photonics.

In this article we describe a new scheme for OAM tomography based on Ref.~\cite{HWeyl} which avoids these problems. In our scheme we measure only a mode qubit, thus simplifying the measurement step. The complexity of the scheme resides in the controlled application of Heisenberg-Weyl operators.

Future implementations of our setup can benefit from reconfigurable SPPs \cite{vSPP-Albero, vSPP-Garcia}. These liquid crystal devices could be used as switchable SPPs in the $X_d^m$ gate.

A possible extension of our scheme is quantum tomography for radial modes. Since sorters for radial modes $r$ have been experimentally demonstrated \cite{ZhouRad, GuRad}, the missing element is the optical equivalent of an SPP for radial modes, $\ket{r} \mapsto \ket{r+1}$. Although adding or subtracting arbitrary units of radial quantum number is an open question, recent developments in this direction are promising \cite{RuffatoSPPr, RuffatoSPPr2}.

\begin{acknowledgments}
The authors acknowledge support from a grant of the Romanian Ministry of Research and Innovation, PCCDI-UEFISCDI, project number PN-III-P1-1.2-PCCDI-2017-0338/79PCCDI/2018, within PNCDI III. R.I.~acknowledges support from PN 19060101/2019-2022.
\end{acknowledgments}

\appendix
\section{Calculations for DQC1}
\label{sec:app:calculations_rho}

Here we calculate the outcome of the circuit in Fig.~\ref{fig:circuit}. After the $H$ and $P_\varphi$ gates, the density matrix of the system is:
\be
\rho'= \frac{1}{2}\left( \ket{0}\bra{0}+ e^{-i\varphi}\ket{0}\bra{1}+ e^{i\varphi}\ket{1}\bra{0}+ \ket{1}\bra{1} \right) \otimes \rho
\ee
\begin{widetext}
We now apply the control gate $C(U)$ and obtain 
\begin{eqnarray}
\label{eq:rho_sec}
\rho''=\frac{1}{2}(
\ket{0}\bra{0}\otimes\rho
+\ket{1}\bra{1}\otimes U\rho U^\dagger
+e^{i\varphi}\ket{1}\bra{0} \otimes U\rho
+e^{-i\varphi}\ket{0}\bra{1}\otimes\rho U^\dagger
)
\end{eqnarray}
The two subsystems (the qubit ancilla and the qudit) are now entangled. After applying the second Hadamard gate $H$ on the qubit, the density matrix becomes
\begin{eqnarray}
\label{eq:rho_third_02}
\rho_{out}=\frac{1}{4}\{
\ket{0}\bra{0}\otimes(\rho+ U\rho U^\dagger
+e^{i\varphi} U\rho
+e^{-i\varphi}\rho U^\dagger)
+\ket{1}\bra{0}\otimes(\rho- U\rho U^\dagger
-e^{i\varphi} U\rho
+e^{-i\varphi}\rho U^\dagger)
\nonumber\\
+\ket{0}\bra{1}\otimes(\rho- U\rho U^\dagger
+e^{i\varphi} U\rho
-e^{-i\varphi}\rho U^\dagger)
+\ket{1}\bra{1}\otimes(\rho+ U\rho U^\dagger
-e^{i\varphi} U\rho
-e^{-i\varphi}\rho U^\dagger)
\}
\end{eqnarray}
In DQC1 we discard the qudit and measure only the qubit ancilla. Hence we trace out the qudit degrees of freedom and we obtain the reduced density matrix $\rho_2$ of the qubit
\begin{eqnarray}
\label{eq:rho_reduced}
\rho_2= \frac{1}{4} (
\ket{0}\bra{0}(2
+\tr\{e^{i\varphi} U\rho
+e^{-i\varphi}\rho U^\dagger\})
+\ket{1}\bra{1}(2-\tr\{e^{i\varphi} U\rho
+e^{-i\varphi}\rho U^\dagger\} )
\nonumber\\
+\ket{1}\bra{0}\tr\{
-e^{i\varphi} U\rho
+e^{-i\varphi}\rho U^\dagger\}
+\ket{0}\bra{1}\tr\{
e^{i\varphi} U\rho
-e^{-i\varphi}\rho U^\dagger\})
\end{eqnarray}
Thus we recover the expectation value of the qubit in the $Z$-basis, eq.~\eqref{ev_Z}:
\be
\langle Z \rangle= \frac{1}{2} \tr\{ \rho (e^{i\varphi} U+ e^{-i\varphi} U^\dag) \}
\ee

\end{widetext}

%
%
\bibliographystyle{apsrev4-1}

\bibliography{Qudit_tomo_bibtex}

\end{document}